\documentstyle[preprint,eqsecnum,aps]{revtex}
\newcommand{\eV}{{\rm eV}}

\begin{document}

\title{Interchain Pair Hopping of Solitons and Polarons via Dopants 
in Polyacetylene}

\author{Makoto Kuwabara, Shuji Abe}
\address{Electrotechnical Laboratory, Tsukuba, Ibaraki 305}
\author{Yoshiyuki Ono}
\address{Department of Physics, Toho University, Funabashi, Chiba 274}

\date{\today}
\maketitle

\begin{abstract}
Interchain hopping of solitons and polarons 
in polyacetylene is studied by numerical simulation of their motion 
under an electric field.  
Use is made of 
the Su-Schrieffer-Heeger model supplemented with 
intrachain electron-electron interactions and dopant potentials.  
We find that charged solitons can hop to the opposite chain by forming 
bound pairs (bipolarons).  
For the case of polarons also, hopping in a pair is more favorable than 
single polaron hopping.  
Interchain hopping of a polaron pair is more efficient than that of a 
soliton pair.   
\end{abstract}


\section{Introduction}
The mechanism of charge hopping among chains is one of the intricate 
unsettled questions related to understanding of the electronic properties 
of conducting polymers.\cite{Polymer_Springer}  
It is expected that charged solitons or polarons play a major role in 
conduction processes in polyacetylene.  
Motion of such nonlinear excitations can contribute directly to 
the conduction.\cite{Conduction,Activation}  
Polyacetylene, however, has a complex structure composed of polymer 
chains with finite lengths and dopant ions (Fig. 1).  
A difficulty in the charged soliton model 
is that a single soliton cannot hop to neighboring chains, 
since there exists a large activation barrier for inversion of the phase 
of bond alternation over a large region.  
The difficulty can be removed by consideration of pairs of solitons, 
i.e. bipolarons.  
Interchain hopping of polarons, which is another conduction mechanism, 
is considered to be much easier.  
It is crucial to consider the dynamics of the nonlinear excitations by 
taking into account the three-dimensional effects, such as dopant potentials 
and interchain interactions.  

Silva and Terai\cite{Geraldon} investigated the effect of the 
interactions between two chains of polyacetylene on the dynamics of solitons,  
and found that the motion of a soliton is suppressed by a confinement 
effect and that 
the amount of the charge transfer between the two chains via charged and 
neutral solitons is quite small.  
The effect of a dopant has been usually incorporated as the Coulomb potential 
for electrons.  
Yamashiro and coworkers pointed out that interchain transfers are 
enhanced by the presence of the valence orbitals of the dopant 
ions.\cite{Yamashiro}  

We study here interchain hopping of solitons and polarons in polyacetylene 
by numerical simulation of their motion under an electric field.  
We consider two neighboring chains that are coupled with each other 
through dopant ions as shown in Fig. 1(b).  
One of these chains contains a pair of solitons or polarons 
in the initial state.  
The effects of a dopant are twofold: the Coulomb potential and enhancement 
of interchain transfer around the dopant.  
In our model, the interchain transfer is assumed to occur in a small region 
around the dopant, and is neglected in the other regions, for simplicity.

The effect of chain ends on the motion of the nonlinear excitations is also 
important.
For discussion of this effect the dopant is located near 
the end of one of the chains, chain a in Fig. 1(b).  
The other chain (chain b) is arranged so that the dopant is located near 
its  midpoint.  
Thus we choose the structure shown in Fig. 1(b), and 
it seems reasonable to regard this structure as a typical example of part 
of the three-dimensional complex structure shown in Fig. 1(a).

\section{Model and Numerical Method}

The model Hamiltonian is given by
\begin{equation}
H = H_{\rm a} + H_{\rm b} + H_{\rm e-e} + H_{\rm int} + H_{\rm imp}. 
\label{eq:totalH}
\end{equation}
The first two terms, describing the two chains a and b, are 
the Su-Schrieffer-Heeger model\cite{SSH} modified to include the 
electric field.  
\begin{eqnarray}
H_\lambda & = & -{\sum_{i\sigma}}
   (t_0-\alpha y_{\lambda, i})
   [{\rm e}^{{\rm i}\gamma A}
     c^\dagger_{\lambda, i\sigma}c_{\lambda, i+1\sigma}+h.c.]
                                                            \nonumber \\
      &   & +\frac{K}{2}{\sum_i}y_{\lambda, i}^2 
            +\frac{M}{2}{\sum_i}{\dot u}_{\lambda, i}^2
            -\Gamma (u_{\lambda, N} - u_{\lambda, 1}), 
\label{eq:SSH}
\end{eqnarray}
where 
$\lambda = {\rm a,b}$ and 
$y_{\lambda, i} = u_{\lambda, i+1}-u_{\lambda, i}$.  
Here, $t_0$ is the hopping integral between the nearest neighbor sites, 
$\alpha$ the electron phonon coupling constant, $u_{\lambda, i}$ 
the displacement of the $i$th CH unit on the chain $\lambda$, 
$c_{\lambda, i\sigma}$ the annihilation 
operator of an electron at the $i$th site with spin $\sigma$ on 
the chain $\lambda$, 
$K$ the spring constant, and $M$ the CH mass. 
The electric field is applied in a direction parallel to the chain axis, 
being given by the time dependent vector potential, 
$E=-\dot{A}/c$.\cite{SdynI}  
The parameter $\gamma$ is defined by $\gamma=ea/\hbar c$, where 
$e$ is the absolute value of the electron charge, $a$ the lattice constant, 
and $c$ the light velocity.  
We adopt the open boundary condition for each chain.  
The last term in eq. (\ref{eq:SSH}) is added to keep the 
chain length constant.  
The third term in eq. (\ref{eq:totalH}) describes the short-range 
intrachain electron-electron interactions.  
\begin{equation}
H_{\rm e-e} =U \sum_{\lambda, i} n_{\lambda, i\uparrow}n_{\lambda, i\downarrow}
            +V \sum_{\lambda, i} n_{\lambda, i} n_{\lambda, i+1},
\end{equation}
with 
$n_{\lambda, i\sigma} = c^\dagger_{\lambda, i\sigma}c_{\lambda ,i\sigma}$ and 
$n_{\lambda, i} = \sum_\sigma n_{\lambda, i\sigma}$.  
The parameters $U$ and $V$ are the on-site and nearest neighbor Coulomb 
repulsion, respectively.  
The two chains are of the same size. 
Chain b is shifted by $i_0$ sites with respect to chain a.  
The interaction between two chains is represented by
\begin{equation}
H_{\rm int} = -t_\perp \sum_{\sigma, i=p}^{q}
                      (c^{\dagger}_{{\rm a},i\sigma} c_{{\rm b},i-i_0\sigma}
                     + c^{\dagger}_{{\rm b},i-i_0\sigma} c_{{\rm a},i\sigma}),
\end{equation}
where $p$ and $q$ denote the first and last site in the interaction region 
around the dopant.  
The last term of eq.(\ref{eq:totalH}) represents the dopant potential, 
which is approximated by a square-well potential,
\begin{equation}
H_{\rm imp} = V_{\rm imp} \sum_{\sigma, i=p}^q 
                  (c^{\dagger}_{{\rm a},i\sigma} c_{{\rm a},i\sigma}
                 + c^{\dagger}_{{\rm b},i-i_0\sigma} c_{{\rm b},i-i_0\sigma} ).
\end{equation}
We use the following values of parameters:\cite{SSH,Yonemitsu}  
$t_0=2.5\eV$, $t_\perp = 0.1t_0$, 
$K=21\eV/{\rm \AA^2}$, 
$\alpha=4.1\eV/{\rm \AA}$, $a=1.22{\rm \AA}$, $V_{\rm imp}=-0.1t_0$, 
$U=t_0$, $V=t_0/2$ and $\Gamma=5.13\eV$.  
The chain sizes are $N_{\rm a}=N_{\rm b}=120$ 
and the total number of electrons is $N_{\rm e}=242$, corresponding 
to two excess electrons in the system.  
The center of the dopant is located at the 105th site in the coordinate of 
chain a.  
The endpoints of the interaction region are set to be $p=100$ and 
$q=110$.  
The shift of chain b with respect to chain a is $i_0=46$, so that 
the dopant is located near the center of chain b.  

The initial states are prepared self-consistently in the Hartree-Fock 
approximation, in which chain a contains two solitons or polarons, 
one of which is pinned by the dopant and the other of which is free to move.  
The time evolution of the electronic wavefunction is determined by use of the 
time-dependent HF equation, and the motion of the lattice 
by use of an equation of motion.\cite{SdynI,Terai}  
In solving the time dependent HF equation we use the method of fractal 
decomposition for exponential operators.\cite{Suzuki,Terai_prog}  
For the equation of motion the time differential equation is integrated 
with discretization of the time with an interval $\Delta t$.  
The interval $\Delta t$ is chosen to be 
$\Delta t = 0.0025\omega_{\rm Q}^{-1}$ throughout this work.  
Here in our simulations the unit of time is 
the inverse of the bare optical phonon frequency 
($\omega_{\rm Q}=\sqrt{4K/M}\approx 2.5\times 10^{-14}{\rm s}^{-1}$)
and that of the electric field is $E_0=\hbar\omega/ea$ 
$(\approx 1.3\times 10^7 {\rm V}/{\rm cm})$.

\section{Results and Discussion}
\subsection{Case of a pair of solitons}
In the following we show the results in terms of the smoothened quantities 
of the bond 
variable and the excess charge distribution, defined as
\begin{eqnarray}
\tilde{y}_{\lambda, i} & = & (-1)^{i}
                (y_{\lambda, i-1} -2y_{\lambda, i} + y_{\lambda, i+1}) /4, \\
\tilde{\rho}_{\lambda, i} & = &
           (\rho_{\lambda, i-1} + 2\rho_{\lambda, i} + \rho_{\lambda, i+1}) /4,
\end{eqnarray}
where the excess charge density is calculated from the time-dependent 
wavefunctions $\{ \Psi_{\lambda \nu}(i,t)\} $, 
\begin{equation}
\rho_{\lambda, i}(t) = {\sum_\nu}^{'} |\Psi_{\lambda \nu}(i,t)|^2 -1.
\end{equation}
The prime attached 
to the summation symbol denotes a sum over occupied states.  

Figure 2 shows a typical example of hopping of a pair of solitons 
from chain a to b in three-dimensional representation of 
$\tilde{y}_{\lambda i}$ and $\tilde{\rho}_{\lambda i}$ as 
functions of time and space.  
The free soliton on chain a 
is accelerated by the field and collides with the pinned 
soliton, resulting in formation of a bipolaronic state.  
In chain b small oscillations are generated after the field 
is switched on.  
Their amplitudes increase upon the collision and sometimes 
the oscillations induce bipolaronic lattice distortion.  
In such a case the soliton pair can easily hop to the neighboring chain.  
After the hopping, one of the solitons is pinned by the dopant and the other 
begins to move toward the chain end.

We have performed numerical simulations for various strengths of the 
field for changing of the incident velocity of 
the free soliton at the 
collision with the pinned soliton.  
In Fig. 3 we display the time dependence of the excess charge density 
on chain a, which is defined by 
\begin{equation}
\rho_{\rm a}(t) = \sum_{i=1}^{N_{\rm a}} \rho_{{\rm a},i}(t).  
\end{equation}
In the initial state $\rho_{\rm a}(0)$ has almost $2|e|$, since there 
exist a pair of charged solitons on chain a.  
A slight deviation from $2|e|$ is due to partial flow 
of charge from the pinned soliton to chain b.  
The deviation increases and fluctuates after the two solitons form 
a bipolaron.  
In the case of a weak field (i.e. low incident velocity at the collision), 
the soliton pair does not show any sign of hopping at least within 
the longest time of our simulations.  
For a stronger field (high incident velocity), $\rho_{\rm a}(t)$ falls 
to about zero suddenly, indicating that the hopping of two electrons occurs 
simultaneously.  
There is a threshold velocity for the hop.  
This means that there is a potential barrier for the hopping.  
The time interval between the collision and completion of the hop 
is not a simple function of the field strength (the incident velocity), 
but a rather chaotic one.  
The occurrence of the hop depends on the mutual motion 
of the soliton pair on chain a and the lattice oscillation on chain b.  
Namely, the hopping occurs under the conditions that 
the amplitude of the bipolaron becomes small on chain a and 
instantaneously becomes large on chain b.  
Therefore a slight difference in parameters results in a drastic 
change in the hopping time.  
Bipolaronic lattice distortion on the opposite chain is necessary for the 
hop to occur.  
The lattice fluctuations at finite temperature may affect 
the hopping, perhaps they make the hop easier.  

The location of the dopant affects the hopping probability.  
If the dopant is located near the midpoint of chain a, 
the probability of hopping of the soliton pair is very low.  
When the field is sufficiently strong, the pinned soliton is pushed out 
by its collision with the free soliton and begins to move 
within chain a.  
In the case that the dopant is located near the chain end, even if the pinned 
soliton is pushed out, it is immediately reflected at the chain end, 
resulting in the formation of a bipolaron again.  
Bipolaron formation is necessary for the hopping to occur.  
Therefore, the hopping probability is larger when the dopant is located near 
the chain end.  

\subsection{Case of a pair of polarons}
For the case of a pair of polarons three-dimensional representations of 
$\tilde{y}_{\lambda i}$ and $\tilde{\rho}_{\lambda i}$ are shown in Fig. 4.  
The hopping process seems 
qualitatively different from that for the soliton pair.  
The polaron pinned by the dopant in the initial state hops back and forth 
between the two chains in the electric field.  
The bipolaronic state observed in the case of the soliton pair 
does not appear as an intermediate state during the hop.  
The intermediate state in this case is such that each chain contains one 
polaron around the dopant.  
After the hopping, one of the two polarons is pinned by the dopant 
over the two chains and the other one is moved 
on chain b by the field.  
For the case of a single polaron, 
hardly any charge transport over the two chains due to interchain hopping 
through the dopant takes place, 
because of trapping by the dopant potential.  
In Fig. 5 we show the time dependence of the excess charge density on chain a 
$\rho_{\rm a}(t)$ for several field strengths.  
Before the collision, 
$\rho_{\rm a}(t)$ oscillates at around $1.5|e|$ 
due to the back-and-forth motion of 
the polaron pinned over the two chains.  
For the case $E=0.01E_0$, two polarons are pinned by the dopant 
after the collision and there exists one polaron per chain.  
Therefore amount of the excess charge density on chain a is $|e|$ on average.  
This case does not contribute to the charge transport from chain a to 
chain b.  
In the stronger field cases the charge transport occurs by the polaron 
hopping.  
Then $\rho_{\rm a}$ decreases by $|e|$ on average.  
The potential barrier for the hopping of a polaron pair is smaller than that 
for the hopping of a soliton pair.  
In contrast to the case for a soliton pair, the polaron pair hops almost 
irrespective of the dopant location relative to the chain end.  
When the two polarons collide, the hopping occurs almost instantaneously.  
Therefore, 
the time of the hop varies monotonically 
with the field strength, not in a chaotic manner.  

It seems that the back-and-forth motion of the pinned polaron before the 
collision is induced mainly by the external field.  
For the parameter set used in our simulations, the energy difference between 
the state of the polaron being almost localized on one chain and the state of 
delocalization over the two chains is quite small.  
Therefore the oscillation can be induced by weak disturbance.  
This is considered to be one of the reasons for the small potential barrier 
for the polaron pair hopping.  

It is worth noting that the final states result in formation of 
a pair of solitons or a pair of polarons depending on whether 
the initial state is a pair of solitons or pair of polarons.  
For a single chain with no electron-electron interaction,  
it has been shown by use of the TLM model\cite{TLM} that a two-polaron state 
is unstable against a two soliton-state.\cite{Onodera}  
This is supported by the results of a numerical study on the dynamics 
of a polaron pair: 
it was observed that two polarons dissociate into a pair of solitons 
after their collision.\cite{Terai_SM}  
In our two-chain system, a two-polaron state, 
not a two-soliton state, 
appears after the collision (hopping) of a pair of polarons.  
One of the reasons for this 
may be that the delocalization of polarons over the two chains 
stabilizes the polaron pair state.

In summary, we have numerically investigated interchain hopping of a pair 
of solitons and a pair of polarons.  
We found that charged solitons can hop to the opposite chain by forming a 
composite (bipolaron) in 
the electric field parallel to the chains.  
The existence of both the dopant and the chain end 
assists the hopping.  
The bipolaronic lattice distortion on the opposite chain 
is necessary for the hop to occur.  
There exists a finite potential barrier for the hopping.  
For the case of polarons 
also, hopping in a pair is more favorable than 
single polaron hopping.  
Interchain hopping of a polaron pair is more efficient than that of a 
soliton pair.

\section*{Acknowledgments}
The authors are grateful to Dr. Akira Terai for useful discussions. 
The numerical calculations were performed at the
Research Information Processing System Center, 
Agency of Industrial Science and Technology.

\def\PR{Phys. Rev. }
\def\PRL{Phys. Rev. Lett. }
\def\SSC{Solid State Commun. }
\def\JPS{J. Phys. Soc. Jpn. }
\def\SM{Synthetic Metals }
\def\PL{Phys. Lett. }

\vspace{1cm}
\noindent
Fig. 1. Schematic illustration of (a) polymer chains with dopants 
and (b) the system studied here.

\vspace{1cm}
\noindent
Fig. 2. Stereographic representation of the site ($i$) and time ($t$) 
dependences of (a) the bond variable and (b) the excess charge density 
for the case of 
a pair of solitons.  
The interchain interaction region is depicted as 
the region between two lines in the bottom $i-t$ plane.

\vspace{1cm}
\noindent
Fig. 3. Time dependence of the excess charge density on chain a for 
a pair of solitons.  

\vspace{1cm}
\noindent
Fig. 4. Stereographic representation of the site ($i$) and time ($t$) 
dependences of (a) the bond variable and (b) the excess charge density 
for the case of a pair of polarons.  

\vspace{1cm}
\noindent
Fig. 5. Time dependence of the excess charge density on chain a for 
a pair of polarons.  


\begin{thebibliography}{99}
\bibitem{Polymer_Springer}
H. G. Kiess (Ed.):
{\it Conjugated Conducting Polymers},
Springer Series in Solid-State Sciences 
{\bf 102} (Springer-Verlag, Berlin, 1992).

\bibitem{Conduction}
M. Kuwabara and Y. Ono:
\JPS {\bf 63} (1994) 1081.

\bibitem{Activation}
M. Kuwabara and Y. Ono:
\JPS {\bf 64} (1995) 2106.

\bibitem{Geraldon}
G. M. Silva and A. Terai: \PR B{\bf 47}(1993) 12568.

\bibitem{Yamashiro}
A. Yamashiro, A. Ikawa and H. Fukutome: 
\SM {\bf 65} (1994) 233.

\bibitem{SSH}
W. P. Su, J. R. Schrieffer and A. J. Heeger: 
\PRL {\bf 42} (1979) 1689; 
\PR B{\bf 22} (1980) 2099.

\bibitem{SdynI}
Y. Ono and A. Terai: 
\JPS {\bf 59} (1990) 2893.

\bibitem{Yonemitsu}
K. Yonemitsu, Y. Ono and Y. Wada: 
\JPS {\bf 57} (1988) 3875.

\bibitem{Terai}
A. Terai: 
in {\it Relaxation in Polymers}, edited by
T. Kobayashi 
(World Scientific Publishing Co., Singapore, 1993), p. 269.

\bibitem{Suzuki}
M. Suzuki:
\PL A{\bf 146} (1990) 319; \PL A{\bf 165} (1992) 387; 
\JPS {\bf 61} (1992) 3015; Proc. Japan Acad. {\bf 69} Ser. B (1993) 161.

\bibitem{Terai_prog}
A. Terai and Y. Ono:
Prog. Theor. Phys. Suppl. No. 113 (1993) 177.

\bibitem{TLM}
H. Takayama, Y. R. Lin-Liu and K. Maki: 
\PR B{\bf 21} (1980) 2388.

\bibitem{Onodera}
Y. Onodera and S. Okuno:
\JPS {\rm 52} (1983) 2478.

\bibitem{Terai_SM}
A. Terai and Y. Ono:
\SM {\bf 69} (1995) 681.

\end{thebibliography}
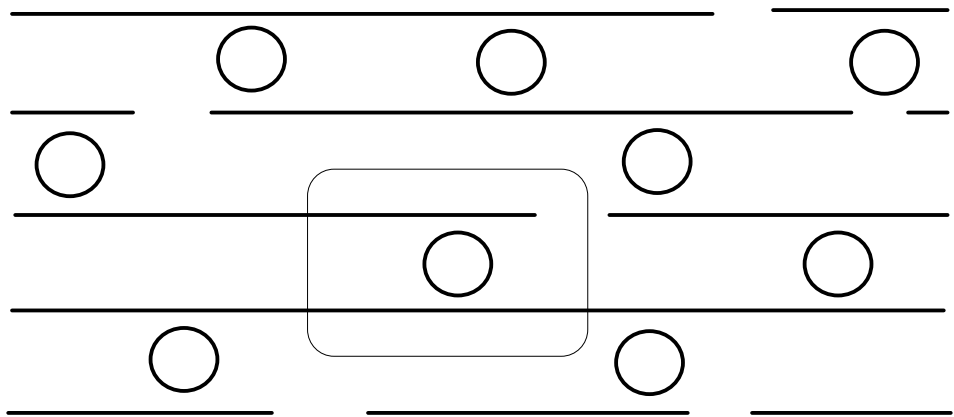
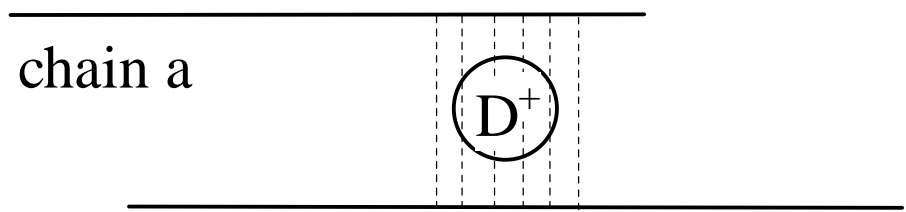
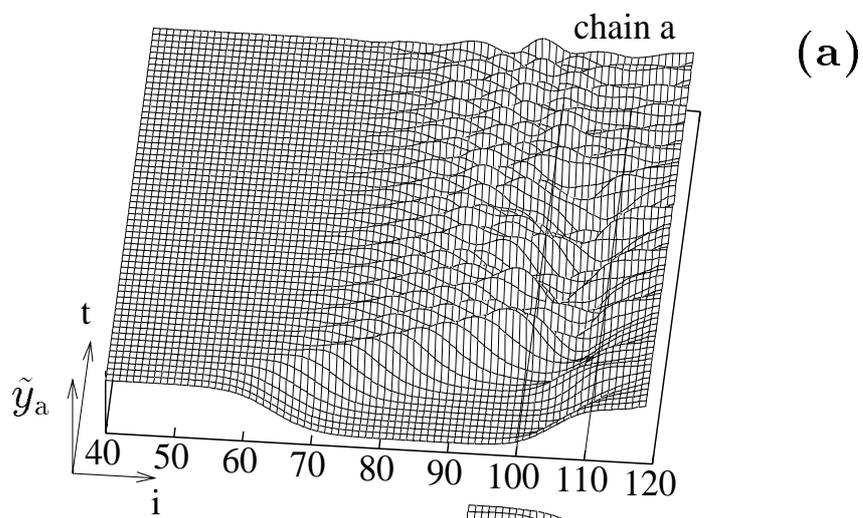
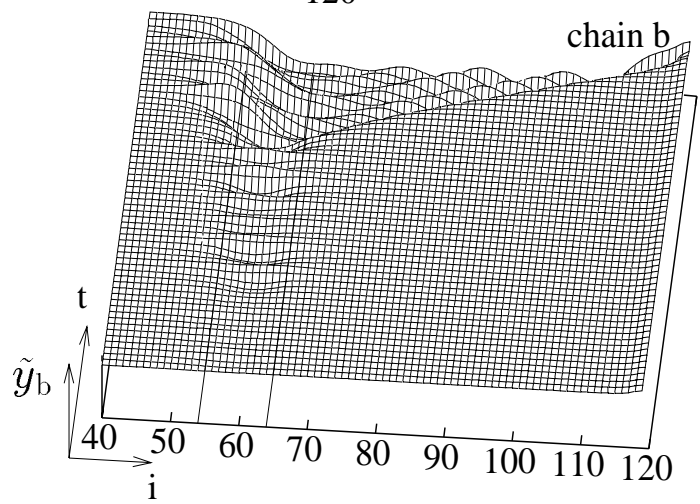
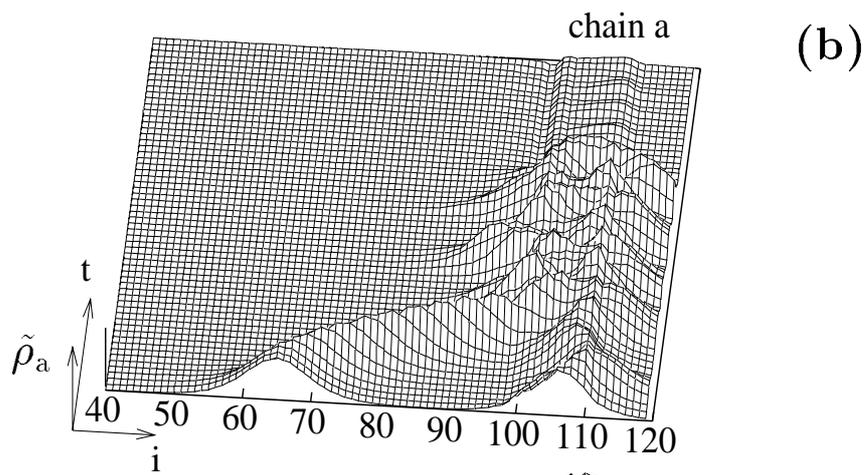
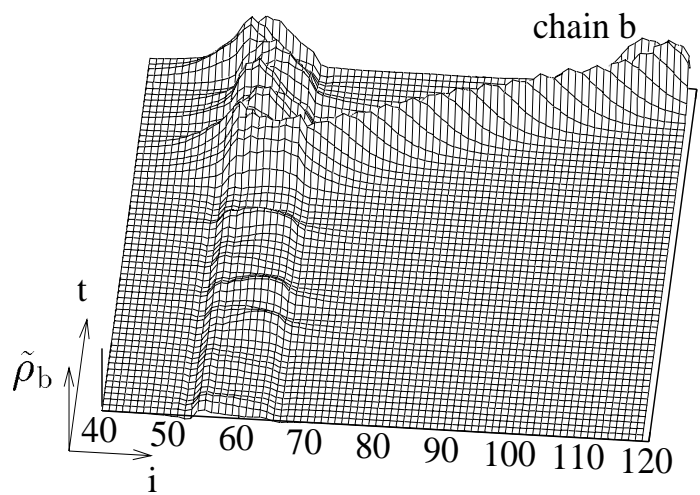
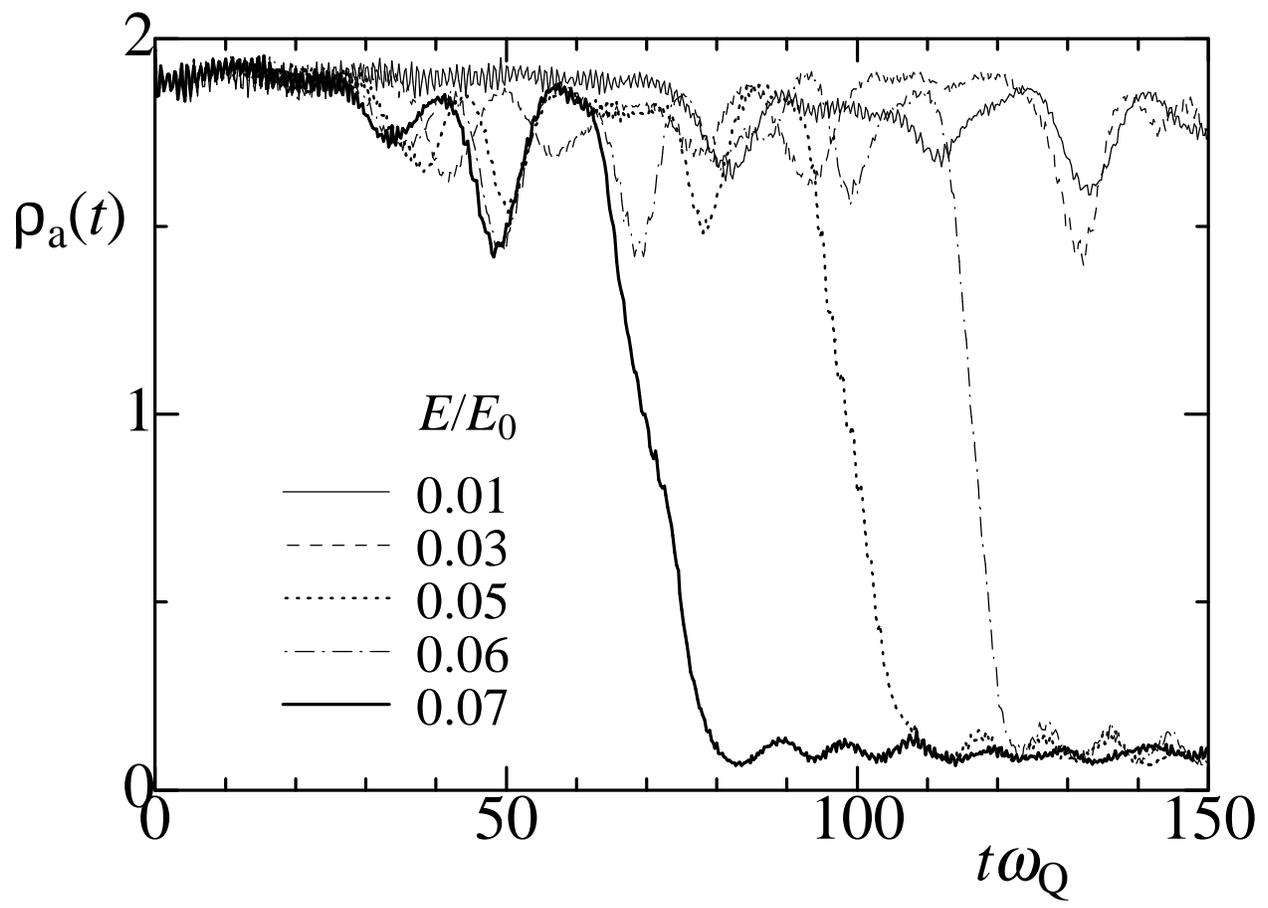
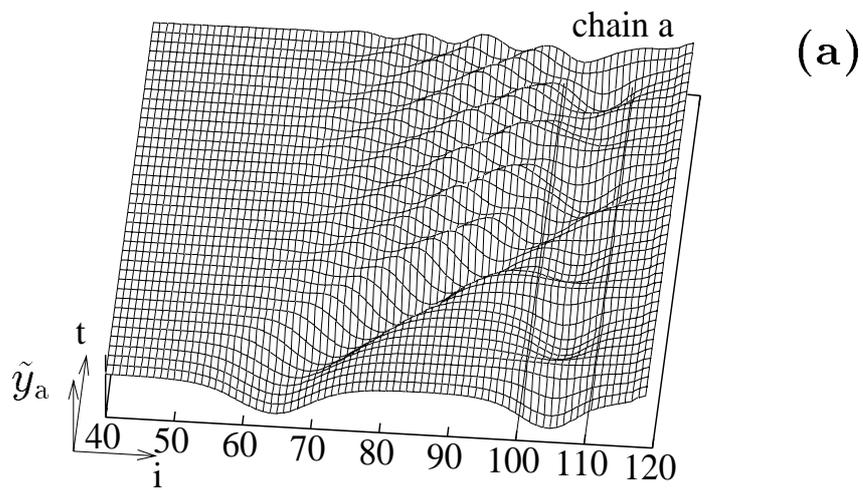
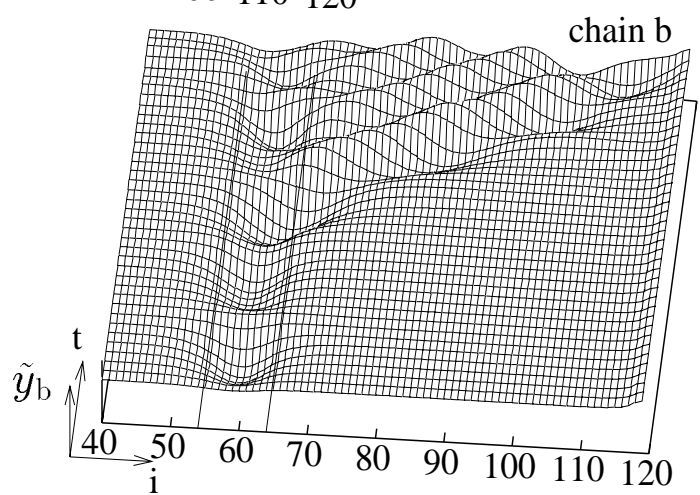
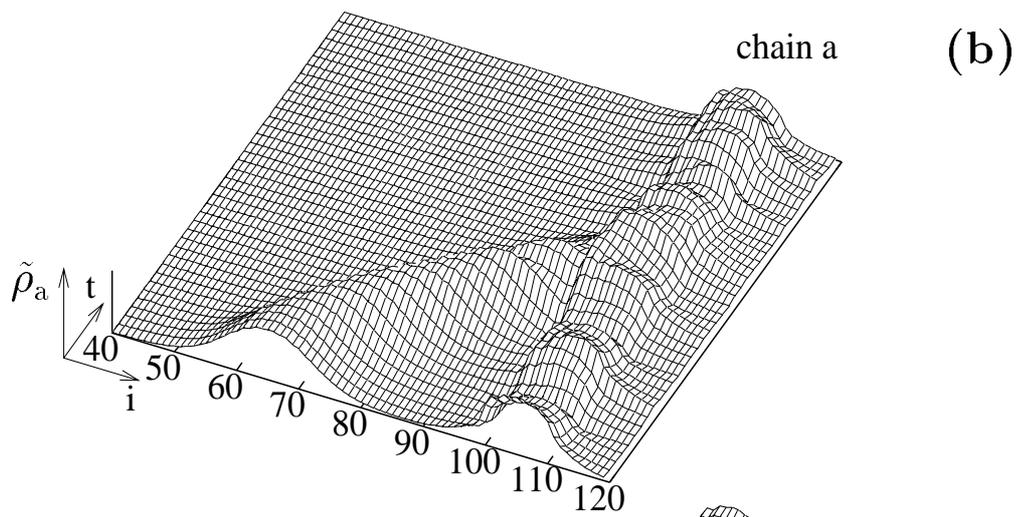
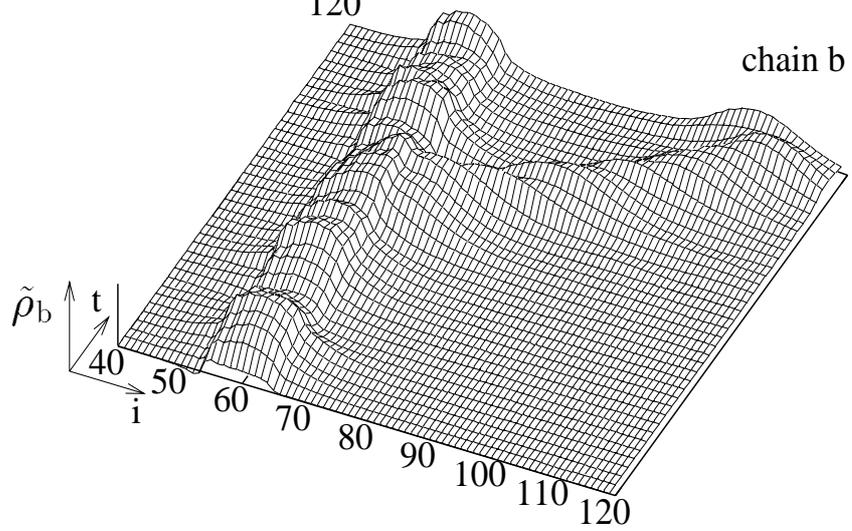
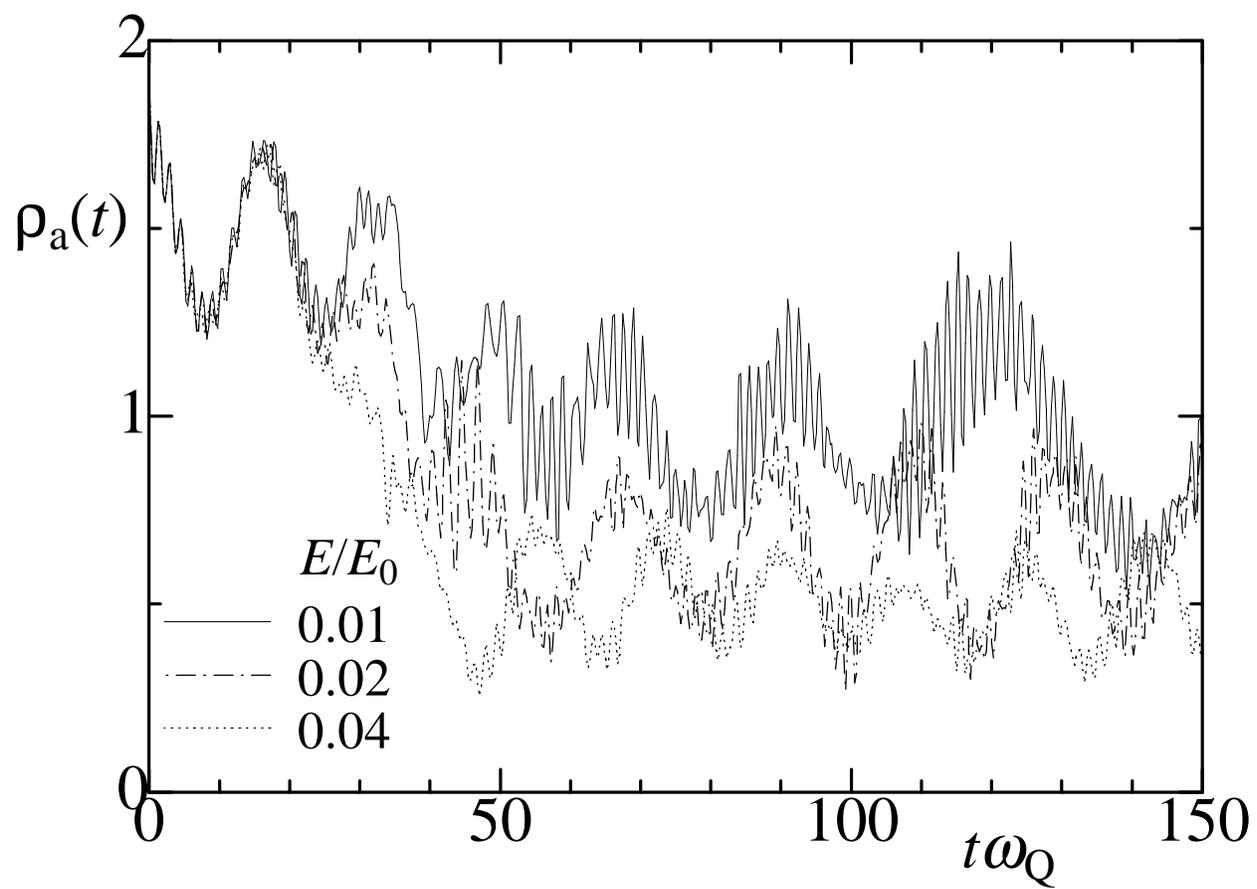
\end{document}